\documentclass[twocolumn,floatfix,preprintnumbers,nofootinbib,superscriptaddress]{revtex4}

\usepackage{ulem}
\usepackage{bm}
\usepackage{times}
\usepackage{amssymb,amsbsy,amsmath,amsfonts}
\usepackage{graphicx}
\usepackage{float}
\usepackage{color}
\usepackage{morefloats}
\usepackage{rotating}
\usepackage{srcltx}
\usepackage{slashed}
\usepackage{subfigure}
\usepackage{multirow}
\usepackage{verbatim}
\usepackage{hyperref}
\usepackage{tabularx}
\usepackage{adjustbox}
\usepackage{array}
\usepackage{tabularray}
\begin{document}
\title{Review of the low-lying excited baryons $\Sigma^*(1/2^-)$}

\author{En Wang}\email{wangen@zzu.edu.cn}
\affiliation{School of Physics, Zhengzhou University, Zhengzhou 450001, China}
\affiliation{Guangxi Key Laboratory of Nuclear Physics and Nuclear Technology, Guangxi Normal University, Guilin 541004, China}\vspace{0.5cm}

%\author{Ying Li}
%\affiliation{School of Physics, Zhengzhou
%	University, Zhengzhou 450001, China}

%\author{Wen-Tao Lyu}
%\affiliation{School of Physics, Zhengzhou
%	University, Zhengzhou 450001, China}

\author{Li-Sheng Geng} \email{lisheng.geng@buaa.edu.cn}
\affiliation{School of Physics, Beihang University, Beijing 102206, China}
\affiliation{Peng Huanwu Collaborative Center for Research and Education, Beihang University, Beijing 100191, China}
\affiliation{Beijing Key Laboratory of Advanced Nuclear Materials and Physics, Beihang University, Beijing 102206, China }
\affiliation{Southern Center for Nuclear-Science Theory (SCNT), Institute of Modern Physics, Chinese Academy of Sciences, Huizhou 516000, China}
\vspace{0.5cm}

\author{Jia-Jun Wu}
\email{wujiajun@ucas.ac.cn}
\affiliation{School of Physics, University of Chinese Academy of Sciences, Beijing 100049, China}
\affiliation{Southern Center for Nuclear-Science Theory (SCNT), Institute of Modern Physics, Chinese Academy of Sciences, Huizhou 516000, China}\vspace{0.5cm}

\author{Ju-Jun Xie} \email{xiejujun@impcas.ac.cn}
\affiliation{Southern Center for Nuclear-Science Theory (SCNT), Institute of Modern Physics, Chinese Academy of Sciences, Huizhou 516000, China}
\affiliation{Institute of Modern Physics, Chinese Academy of Sciences, Lanzhou 730000, China} \affiliation{School of Nuclear Sciences and Technology, University of Chinese Academy of Sciences, Beijing 101408, China} 
\vspace{0.5cm}

\author{Bing-Song Zou} \email{zoubs@mail.tsinghua.edu.cn}
\affiliation{Department of Physics, Tsinghua University, Beijing 100084, China}
\affiliation{CAS Key Laboratory of Theoretical Physics, Institute of Theoretical Physics,
Chinese Academy of Sciences, Beijing 100190,China} \affiliation{School of Physics, University of Chinese Academy of Sciences, Beijing 100049, China} 
\affiliation{Southern Center for Nuclear-Science Theory (SCNT), Institute of Modern Physics, Chinese Academy of Sciences, Huizhou 516000, China}
\vspace{0.5cm}

\begin{abstract}

Strong empirical and phenomenological indications exist for large sea-quark admixtures in the low-lying excited baryons. Investigating the low-lying excited baryon $\Sigma^*(1/2^-)$ is important to determine the nature of the low-lying excited baryons. We review the experimental and theoretical progress on the studies of the $\Sigma^*(1/2^-)$. 
Although several candidates have received intensive discussions, such as $\Sigma(1620)$ and $\Sigma(1480)$, their existence needs further confirmation. 
Following the prediction of the unquenched quark models for the $\Sigma^*(1/2^-)$, many theoretical works suggested the existence of these states in various processes. 
Future experimental measurements could shed light on the existence of the low-lying excited $\Sigma^*(1/2^-)$ state.
\end{abstract}

%\keywords{ }
\date{\today}

\maketitle

\section{Introduction}\label{sec1}
As composite subatomic particles made of quarks and gluons held together by the strong interaction, hadrons are categorized into $q\bar{q}$ mesons and $qqq$ baryons within the conventional quark model proposed by Gell-Mann and Zweig~\cite{Gell-Mann:1964ewy,Zweig:1964ruk}. 
So far, all the ground-state mesons and baryons composed of light quarks ($u,d,s$) have been well established, and a huge number of excited light mesons and baryons have been observed experimentally~\cite{ParticleDataGroup:2022pth}, some of which can not be well described in the conventional quark models.
For instance, recently BESIII has observed one isoscalar state $\eta_1(1855)$ with exotic quantum numbers $J^{PC}=1^{-+}$ in the process $J/\psi \to \gamma \eta \eta'$, which could be a candidate for hybrid hadrons~\cite{BESIII:2022riz,Chen:2022qpd}. 
Those exotic hadrons challenge our understanding of the quark model and provide insights into the strong interaction dynamics within Quantum Chromodynamics (QCD)~\cite{Chen:2016spr,Chen:2022asf,Guo:2017jvc,Oset:2016lyh,Liu:2024uxn,Gao:1999ar,Hall:2014uca,Zhang:2024ery,Shi:2023ntq,Wang:2023vtx,Wang:2022xga,Xie:2024wbd}.

\begin{figure}[htbp]
		\centering
		%	\centering
		\includegraphics[scale=0.45]{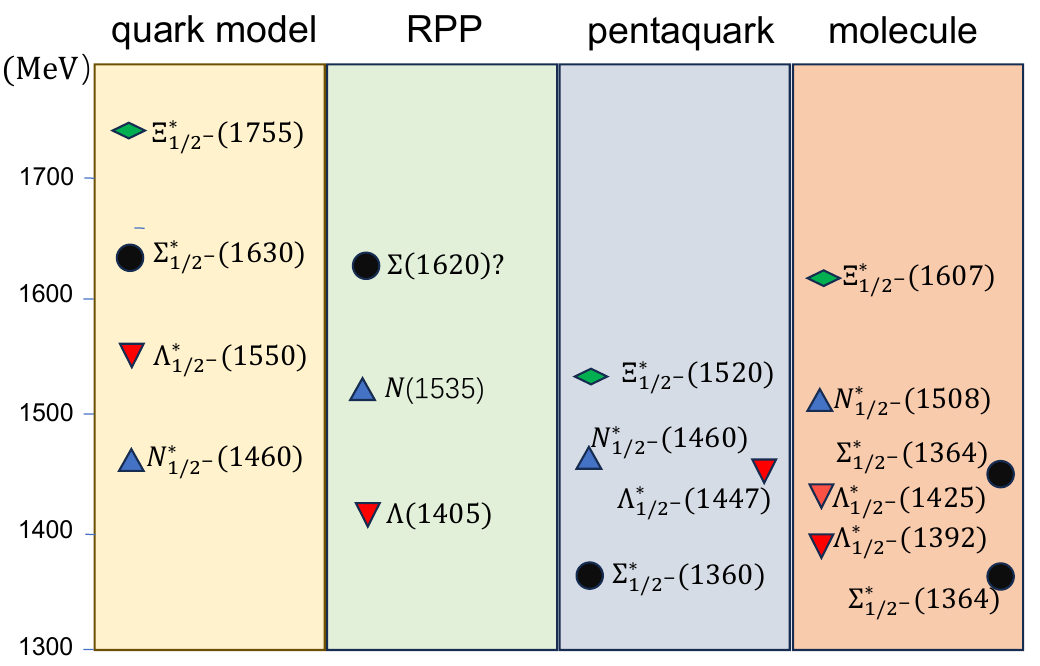}
	\caption{Masses of the low-lying excited baryons with $J^P=1/2^-$. The results of `quark model' are taken from the predictions of the quark model~\cite{Capstick:1986ter}, the results of `pentaquark' are taken from the predictions of the pentaquark model~\cite{Zhang:2004xt}, and the results of `RPP' are the experimental data~\cite{ParticleDataGroup:2022pth}. In addition, the $N^*_{1/2^-}$, $\Lambda^*_{1/2^-}$/$\Sigma^*_{1/2^-}$, and $\Xi^*_{1/2^-}$ masses of `molecule' are taken from the predictions of Ref.~\cite{Garzon:2014ida}, Ref.~\cite{Lu:2022hwm}, and Ref.~\cite{Ramos:2002xh}, respectively.} \label{fig:mass}
\end{figure}

There are some puzzles for the low-lying excited baryons with the quantum numbers of spin-parity $J^P=1/2^-$. 
One puzzle is the ``mass reverse problem''. 
In the naive quark model, the mass of the $N(1535)$ with spin-parity quantum numbers $J^P=1/2^-$ should be lower than the one of the radial excitation $N(1440)$ with $J^P=1/2^+$, and the one of the $\Lambda(1405)$ ($J^P=1/2^-$) with strangeness $S=-1$~\cite{ParticleDataGroup:2022pth,Liu:2005pm,Geng:2008cv,Capstick:1986ter}, as shown in Fig.~\ref{fig:mass}.
%
%\sout{\textcolor{red}{Besides, the $N(1535)$ resonance has also other debatable problems. It is found in Ref.~\cite{Liu:2005pm} that this resonance couples stronger to the $K\Lambda$ channel, rather than $\eta N$ channel, which has been challenged by the study of Ref.~\cite{Mart:2013ida}. In Ref.~\cite{Mart:2013ida}, the coupling of the $N(1535)$ to the $K^+\Lambda$, that extracted from the largest $K^+\Lambda$ photoproduction database by using an isobar model, is smaller than the results of Ref.~\cite{Liu:2005pm}.
%In a word, the $N(1535)$ strongly couples to the channels with strangeness, which implies that this state has significant sea-quark admixtures.}}
%
Although there are some debates on the relative ratio between the coupling constants of $N(1535)N\eta$ and $N(1535)K\Lambda$~\cite{Liu:2005pm,Geng:2008cv,Mart:2013ida,Wang:2022osj} obtained in different ways, they all
indicate that the $N(1535)$ strongly couples to the channels with strangeness, which implies that this state has a significant penta-quark component with hidden strangeness.
Concerning another low-lying $J^P=1/2^-$ excited baryon $\Lambda(1670)$, it exhibits itself as an enhancement in the $\eta\Lambda$ invariant mass distribution of the processes $K^-p\to \eta\Lambda$~\cite{CrystalBall:2001uhc} and $\Lambda_c^+\to \eta\Lambda\pi^+$~\cite{BESIII:2018qyg,Belle:2020xku}, and as a dip structure around the $\eta\Lambda$ threshold in the $\bar{K}^0n$ invariant mass distribution of the process $K^-p\to \bar{K}^0n$~\cite{Gopal:1976gs}. Recently, the Belle Collaboration observed a cusp structure around the $\eta\Lambda$ threshold in the $pK^-$ invariant mass distribution of the process $\Lambda_c^+\to pK^-\pi^+$, which could be associated with the $\Lambda(1670)$~\cite{Belle:2022cbs,Zhang:2024jby}.
Thus, these exotic properties of the low-lying excited baryons with the quantum numbers of spin-parity $J^P=1/2^-$ are difficult to explain in the simple quenched quark model and should have a much richer structure than the three valence quark composition~\cite{Zou:2010tc}.

\begin{figure}[htbp]
		\centering
		%	\centering
		\includegraphics[scale=0.4]{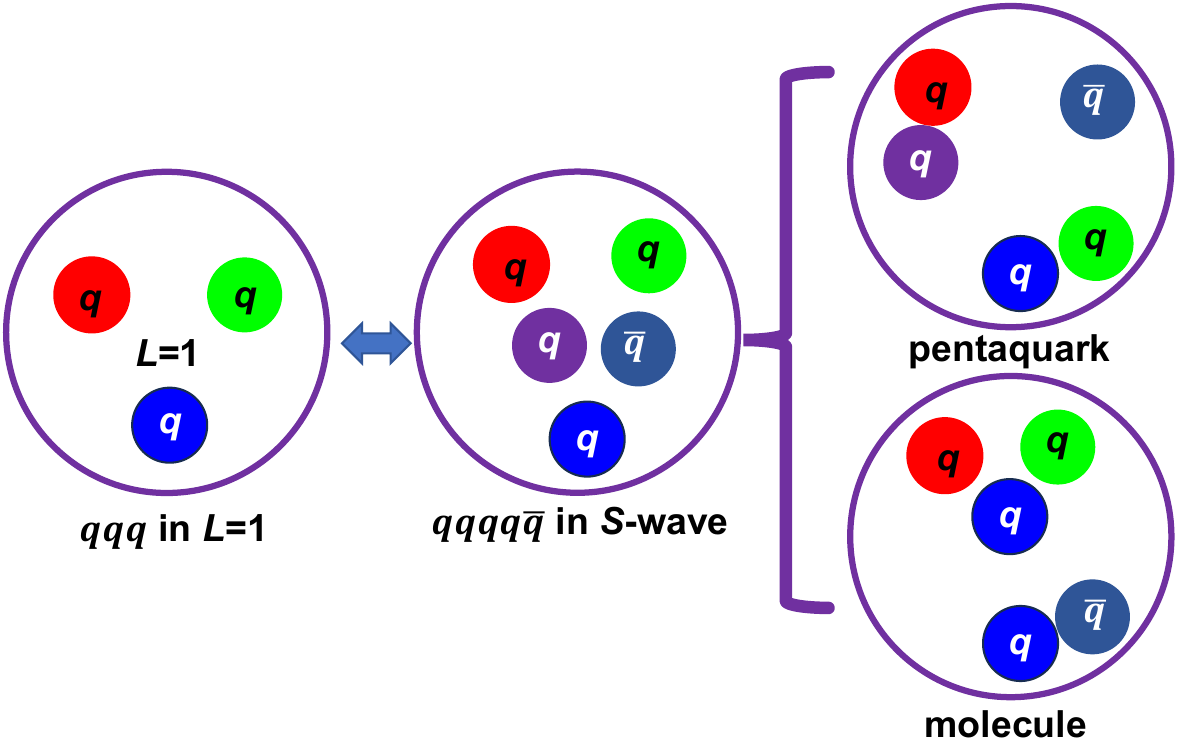}
	\caption{The $qqq$ and $qqqq\bar{q}$ configurations for the low-lying excited baryons with $J^P=1/2^-$. } \label{fig:pentaquark}
\end{figure}

In the classical quenched quark model, which ignores the creation of the quark-antiquark pairs, the quarks cannot be separated from a hadron due to an infinitely large confinement potential. However, as shown in Fig.~\ref{fig:pentaquark}, the spatial excitation energy of a quark in a baryon is comparable to pulling a $q\bar{q}$ pair from the gluon field~\cite{Zou:2016bxw}, which implies that the creation of quark-antiquark pairs from the vacuum plays a crucial role in understanding quark confinement and hadron spectroscopy.
The 5-quark components in baryons can be either in the form of molecule, such as $\bar{K}N-\Sigma\pi$ for $\Lambda(1405)$, or in other forms of quark correlations, such as pentaquark configurations $[ud][us]\bar{s}$ for $N(1535)$, as depicted in Fig.~\ref{fig:pentaquark}. 

The low-lying baryon $\Lambda(1405)$ with $J^P=1/2^-$ is difficult to understand in the naive quark model as a $P$-wave excited baryon~\cite{Isgur:1978xj}. It has been proposed that this state can be generated by the $\Bar{K}N$ coupled-channel interaction within the unitary chiral approach~\cite{Kaiser:1995eg,Oset:1997it,Oller:2000fj,Jido:2003cb}, or contains both a $qqq$ component and a $qqqq\bar q$ component that is strongly coupled to the $\bar K N$ and $\pi\Sigma$ channels~\cite{Helminen:2000jb}.
%
%Besides the strong couplings of $N(1535)$ to the $\eta N$ and $K \Lambda$ channels, there is also evidence for large effective $g_{N^(1535) N \eta'}$ coupling from $\gamma p \to p \eta'$ reaction at CLAS~\cite{CLAS:2005rxe} and $pp \to pp \eta'$ reaction~\cite{Cao:2008st}.
%
In Refs.~\cite{Helminen:2000jb,Zou:2007mk,Zhang:2004xt}, it was proposed that the $N(1535)$ resonance could be the lowest $L=1$ orbital excited $uud$ state with a large admixing of the $[ud][us]\bar{s}$ pentaquark component, which makes $N(1535)$ heavier than $N(1440)$, and provides a natural explanation for the large couplings of $N(1535)$ to the strangeness channels $\eta N$, $K\Lambda$, and $K\Sigma$. 
In addition, $N(1535)$ could be dynamically generated from the $S$-wave pseudoscalar meson-octet baryon interactions within the chiral unitary approach and is predicted to strongly couple to the channels $\eta N$, $K\Lambda$, and $K\Sigma$~\cite{Kaiser:1995eg,Kaiser:1996js,Inoue:2001ip,Bruns:2010sv,Nieves:2011gb, Gamermann:2011mq, Khemchandani:2013nma}. 
The physical picture remains unchanged, even after the pseudoscalar meson-baryon mixing with the vector-baryon sates was considered~\cite{Garzon:2014ida}. 
The $\phi$ production in the $\pi p$ and $pp$ collisions suggest that the $N(1535)$ strongly couples to the $K\Sigma$, $K\Lambda$, and $\phi N$ channels~\cite{Xie:2007qt,Doring:2008sv}, consistent with the results of the chiral unitary approach.
We have proposed to study the $N(1535)$ within the chiral unitary approach through the three-body decays of $\Lambda_b$~\cite{Lyu:2023aqn,Lu:2016roh,Wang:2015pcn}. 
Recently, it was suggested to test the molecular nature of $N(1535)$ by measuring its correlation functions~\cite{Molina:2023jov}, or the scattering length and effective range of the  $K\Sigma$, $K\Lambda$, and $\eta p$ channels~\cite{Li:2023pjx}.
The $N(1535)$ was also interpreted as the three-quark core dressed by meson-nucleon scattering contributions within the Hamiltonian Effective Field Theory~\cite{Abell:2023nex,Guo:2022hud,Liu:2015ktc}.

%
%

%
%
%and another low-lying baryon $N(1535)$ could also be explained as a dynamically generated state via the pseudoscalar meson-octet baryon interactions~\cite{Inoue:2001ip}.
%{\color{red} for N(1535), it is not so clear, and I guess it will be mixing with 3q state a lot, may we do not mention this state?}
%
However, the low-lying excited $\Sigma^*(1/2^-)$ has not yet been well established experimentally and theoretically~\cite{ParticleDataGroup:2022pth}.
Although the mass of the $\Sigma(1620)$ with $J^P=1/2^-$ listed in the Review of Particle Physics (RPP)~\cite{ParticleDataGroup:2022pth} is consistent with the quenched quark model's prediction of the $\Sigma^*(1/2^-)$ mass~\cite{Capstick:1986ter,Capstick:2000qj,Glozman:1995fu}, as shown in Fig.~\ref{fig:mass}, its status is only 1-star (meaning that the evidence of its existence is poor). It is omitted from the summary tables, which implies that it still needs to be confirmed.

It should be pointed out that the unquenched quark models have also made some interesting predictions for the isovector partner $\Sigma^*(1/2^-)$ of the $\Lambda(1405)$ and $N(1535)$. 
In Ref.~\cite{Helminen:2000jb}, the authors suggested that most of the low-lying baryon resonances in all flavor sectors of the baryon spectrum have strong sea-quark components of the form $qqqq\bar{q}$, and predicted a $\Sigma^*(1/2^-)$ resonance with a mass about 1509~MeV. 
Within Jaffe and Wilczek's diquark model framework, Ref.~\cite{Zhang:2004xt} predicted the mass of  $\Sigma^*(1/2^-)$ to be about 1360~MeV, as shown in Fig.~\ref{fig:mass}.

On the other hand, the $S$-wave meson-baryon interaction in the $S=-1$ sector was studied using the chiral unitary approach, and one $\Sigma^*(1/2^-)$ with a mass around the $\Bar{K}N$ threshold was predicted in Refs.~\cite{Oset:2001cn,Oller:2000fj,Oset:1997it,Khemchandani:2018amu,Kamiya:2016jqc,Jido:2003cb,Oller:2006jw,Garcia-Recio:2002yxy,Lutz:2001yb,Guo:2012vv}, which is supported by the analysis of the CLAS data on the process $\gamma p \to K\Sigma\pi$~\cite{CLAS:2013rjt,Roca:2013cca}. 

In addition, as shown in Fig.~\ref{fig:mass}, two $\Sigma^*(1/2^-)$ states were predicted  in the covariant baryon chiral perturbation theory, including the next-to-next-to-leading order contributions, with the narrower pole located at $(1432,-i18)$~MeV showing up as a cusp, and the broader one located at $(1364,-i110)$~MeV becoming a broad enhancement in the real axis of the amplitude squared~\cite{Lu:2022hwm}. 
Indeed, Ref.~\cite{Jido:2003cb} also found two poles of $(1401,40i)$~MeV and  $(1488,114)$~Mev within the chiral unitary approach by considering the results obtained from the model of Ref.~\cite{Oller:2000fj}.

In a word, the $\Sigma^*(1/2^-)$ with a mass around 1.4~GeV as a partner of $\Lambda(1405)$ has not been found yet. 
Since searching for the state $\Sigma^*(1/2^-)$ is crucial to deepening our understanding of low-lying excited baryons~\cite{Crede:2013kia,Klempt:2009pi,Oset:2016lyh,Chen:2022zgm,Ramos:2002xh,Azimov:2003bb,Azimov:1970ei} and there are many theoretical works have been devoted to this issue, we would like to review the experimental and theoretical progress on the studies of the low-lying excited baryon $\Sigma^*(1/2^-)$.
 
In this review, we give the experimental progress and theoretical studies about the $\Sigma^*(1/2^-)$ in Section~\ref{sec:exp} and Section~\ref{sec:theo}, respectively. 
Finally, we give a summary and outlook in Section~\ref{sec:sum}.

\section{Experimental information on $\Sigma^*(1/2^-)$}\label{sec:exp}

\subsection{$\Sigma(1620)$}
The $\Sigma(1620)$ with $J^P=1/2^-$ was listed as a 2-star resonance in the previous version of PRR and downgraded to 1-star in the present version~\cite{ParticleDataGroup:2022pth}. 
Although its mass $(1600\sim1650)$~MeV is consistent with the predicted mass around \textcolor{red}{1650~MeV} of the quenched quark models~\cite{Capstick:1986ter}, the evidence for its existence is very weak~\cite{Zou:2016bxw}. 
Based on the multi-channel analysis of the $\bar{K}N$ reactions, both Ref.~\cite{Kim:1971zxa} and Ref.~\cite{Langbein:1972uhb} claimed the evidence for a $\Sigma^*(1/2^-)$ with a mass around 1620~MeV, but predicted different branching ratios for this resonance. 
However, Ref.~\cite{Baillon:1975sk} found no signal for $\Sigma^*(1/2^-)$ between 1600~MeV and 1650~MeV through the analysis of the reaction $\bar{K}N\to \Lambda \pi$, and suggested the existence of the $\Sigma(1660)$ with $J^P=1/2^-$, supported by the multi-channel analysis of the $\bar{K}N$ reactions~\cite{Kim:1971zxa,Langbein:1972uhb}. 
In Ref.~\cite{Carroll:1976gs}, the analysis of the total cross sections for $K^-p$ and $K^-n$ with all proper final states indicated some $\Sigma$ resonances near 1600~MeV with quantum numbers undetermined.  
In Ref.~\cite{Morris:1978ia}, the authors analyzed the reaction $K^-n\to \pi^-\Lambda$ and obtained two solutions, with one solution indicating a $\Sigma^*(1/2^-)$ near 1600~MeV, and the other showing no resonant structure below the $\Sigma(1670)$ ($J^P=3/2^-$).

In order to clarify the status of the $\Sigma(1620)$ and the $\Sigma(1660)$, Ref.~\cite{Gao:2010ve} 
analyzed the differential cross sections and $\Lambda$ polarization for both $K^-p \to \pi^0\Lambda$ 
and $K^-n\to \pi^-\Lambda$ reactions with an effective Lagrangian approach, and 
claimed the evidence for the $\Sigma(1620)$, and found that the Crystal
Ball $\Lambda$ polarization data demand the existence of a $\Sigma$ resonance with $J^P = 1/2^+$ and mass near 1635~MeV, compatible with $\Sigma(1660)$ listed in RPP~\cite{ParticleDataGroup:2022pth}, while the  data do not demand the $\Sigma(1620)$.

Recently, Ref.~\cite{Sarantsev:2019xxm} performed the fit of nearly all published data on medium-energy $K^-p$ elastic, charge exchange, and inelastic scattering and did not find the evidence for the $\Sigma(1620)$. Thus, the existence of the $\Sigma(1620)$ is still in debate and needs to be confirmed with more experimental measurements.

\subsection{$\Sigma(1480)$}

In the 2019 version of the RPP~\cite{ParticleDataGroup:2018ovx}, there is one $\Sigma(1480)$  with a mass of about 1480~MeV and a width of $30\sim 60$~MeV, which has only 1-star, and is omitted from the summary tables. 
This state was first observed in the $\Lambda\pi$ and $\Sigma\pi$ spectra in the reaction $\pi^+p\to (Y\pi)K^+ $~\cite{Pan:1970ez}.
Later, Ref.~\cite{AMSTERDAM-CERN-NIJMEGEN-OXFORD:1980gfp} performed a multichannel analysis of $K^-p\to p\bar{K}^0 \pi^-$ at $4.2$~GeV and observed a 3.5 standard deviation signal at 1480~Mev in the $p\bar{K}^0$ which cannot be explained as a reflection of any competing channel.

The ZEUS Collaboration performed a resonance search in the $K_S^0 p$ and $K_S^0 \bar{p}$ invariant mass spectrum of the inclusive deep inelastic scattering at an $ep$ center-of-mass energy of $300\sim318$~GeV and found one resonance with a mass of $1465.1 \pm 2.9$~MeV and a width of $15.5\pm 3.4$~MeV~\cite{ZEUS:2004lje}.

In addition, one hyperon state $Y^{0*}$ with mass of $1480\pm 15$~MeV and width of $60\pm 15$~MeV was observed in the reaction $pp\to pK^+ Y^{0*}$ with at least 4.5 standard deviations by the ANKE spectrometer at COSY-J$\ddot{\rm u}$lich at $p_{\rm bean}=3.65$~GeV~\cite{Zychor:2005sj}. 
Since the isospin of the $Y^{0*}$ has not been determined, it could either be  the $\Sigma(1480)$, or, alternatively, a $\Lambda$ hyperon. 

However, no more precise data supports the existence of the $\Sigma(1480)$, and the present RPP has removed this state~\cite{ParticleDataGroup:2022pth}.

\subsection{$\Sigma^*(1/2^-)$-like structure}

Very recently, the Belle Collaboration reported the $\Lambda\pi^+$ and $\Lambda\pi^-$ invariant mass distributions of the process $\Lambda_c^+ \rightarrow \Lambda \pi^+ \pi^+ \pi^-$, which show a significant cusp signal around the $\bar{K}N$ threshold~\cite{Belle:2022ywa}. 
When interpreted as a resonance, they found for the $\Lambda\pi^+$ combination a mass of $1434.3\pm0.6$(stat.)$^{+0.9}_{-0.0}$(syst.)~MeV, and an intrinsic width of $11.5\pm2.8$(stat.)$^{+0.1}_{-5.3}$(syst.) MeV with a significance of 7.5$\sigma$, and for the $\Lambda\pi^-$ combination a mass of $1438.5\pm0.9$(stat.)$^{+0.2}_{-2.5}$(syst.)~MeV, and an intrinsic width of $33.0\pm7.5$(stat.)$^{+0.1}_{-23.6}$(syst.)~MeV with a significance of 6.2$\sigma$~\cite{Belle:2022ywa}. 
This state is consistent with the theoretical predicted $\Sigma^*(1/2^-)$ with a mass around $\bar K N$ threshold, which is dynamically generated from the $S$-wave meson-baryon interaction in the $S=-1$ sector within the chiral unitary  approach~\cite{Oset:2001cn,Oller:2000fj,Oset:1997it,Khemchandani:2018amu,Kamiya:2016jqc,Jido:2003cb,Oller:2006jw,Garcia-Recio:2002yxy,Lutz:2001yb}.

\section{Theoretical studies of $\Sigma^*(1/2^-)$}\label{sec:theo}
\subsection{$K^-p\to \Lambda\pi^+\pi^-$, $\Lambda p \to \Lambda p \pi^0$}

In Ref.~\cite{Wu:2009tu}, the authors re-examined some old data of the $K^-p \to \Lambda\pi^+\pi^-$ reactions obtained from an exposure of the Laboratory's 72-in. hydrogen bubble chamber to a separated $K^-$ beam of the Bevatron~\cite{Huwe:1969te}. They found that, besides the well-established $\Sigma^*(1385)$ ($J^P=3/2^+$), there is indeed some evidence for the possible existence of a new $\Sigma^*(1/2^-)$ resonance with the same mass but a broader width. 
Later, the authors re-studied the $K^-p \to \Lambda\pi^+\pi^-$ reaction within the effective Lagrangian approach~\cite{Wu:2009nw}. By comparing with the experimental data of the Lawrence Berkeley Laboratory 25-inch hydrogen bubble chamber, they found strong evidence of the resonance $\Sigma^*(1/2^-)$ with a mass around 1380~MeV.

The experimental measurements of the process $\Lambda p\to \Lambda p\pi^0$ show a strong threshold enhancement in the $\pi^0\Lambda$ invariant mass distribution~\cite{Kadyk:1971tc}. 
Since the $\Sigma(1385)$ with spin-parity $J^P=3/2^+$ decays to $\pi\Lambda$ in $P$-wave, its contribution would be suppressed at low energies. 
Thus, Ref.~\cite{Xie:2014zga} studied this process by considering the proposed $\Sigma^*(1/2^-)$ state, which decays into $\pi\Lambda$ in $S$-wave, and described the near-threshold enhancement fairly well, which supports the existence of the $\Sigma^*(1/2^-)$ state in this process.

\subsection{$\Sigma^*(1/2^-)$ photo- and $\nu$-productions}

In Ref.~\cite{Chen:2013vxa}, the authors studied the reactions $\gamma N \to K^+\Sigma(1385)\to K^+\pi^0\Lambda$ near threshold within an effective Lagrangian approach. They found that the $\Sigma^*(1/2^-)$ and the interference of $\Sigma(1385)$ and $\Sigma^*(1/2^-)$ play significant roles near threshold. 
The predicted ratios of the helicity cross sections $\sigma_{3/2}/\sigma_{1/2}$ and the angular distribution of the $\pi$ are distinctly different depending on whether the $\Sigma^*(1/2^-)$ is present.

In Ref.~\cite{Roca:2013av}, the authors analyzed the CLAS data for the $\Lambda(1405)$ photoproduction in the process $\gamma p \to K^+ \pi^0\Sigma^0$~\cite{CLAS:2013rjt}, which filters isospin $I=0$ for the $\pi^0\Sigma^0$ system, and determined the mass and width of the two $\Lambda(1405)$ resonances, which are dynamically generated from the meson-baryon interactions within the chiral unitary approach. 
Following the same strategy, they extended the method to analyze the experimental data on $\gamma p\to K^+ \pi^\pm \Sigma^\mp $ by incorporating the isospin $I=1$ component within the chiral unitary approach~\cite{CLAS:2013rjt}. They suggested the existence of an $I=1$ resonance around the $\bar{K}N$ threshold~\cite{Roca:2013cca}. 
Even though there is no pole in the usual unphysical Riemann sheet connected to the physical one, the $I=1$ amplitude has a resonance-like structure around the $\bar K N$ threshold, with a clear cusp similar to the case of $a_0(980)$~\cite{Oller:1998hw,BESIII:2016tqo,Liang:2016hmr}.

The LEPS Collaboration measured the $\gamma n \to K^+\Sigma(1385)^0$ reaction using a linearly polarized photon beam and found that the azimuthal angle $\phi$-dependence of the asymmetry is negative at $E_{\gamma}=1.8-2.4$~GeV~\cite{LEPS:2008azb}, in contrast to the theoretical prediction~\cite{Oh:2007jd}. 
By including the proposed $\Sigma^*(1/2^-)$ with a mass around 1380~MeV, Ref.~\cite{Gao:2010hy} well reproduced the experimental data for both the $\gamma n$ and $\gamma p$ experiments.

Based on the cross-section data of $K^-p$ scattering,  Ref.~\cite{Khemchandani:2018amu} studied the hyperon resonances by including both pseudoscalar-baryon and vector-baryon dynamics, and produced the properties of the $\Lambda(1405)$, supporting the existence of an isospin $I=1$ state $\Sigma^*(1/2)$ with a mass around 1400~MeV. Furthermore, they calculated the total and differential cross-sections for $\gamma p \to K^+\Lambda(1405)$, in agreement with the CLAS data~\cite{CLAS:2013rxx}, and predicted the cross-section of the process $\gamma p\to K^+\Sigma^*(1/2^-)$~\cite{Kim:2021wov}.

Recently, Ref.~\cite{Lyu:2023oqn} studied the $\Sigma^*(1/2^-)$ photoproduction within the Regge-effective Lagrangian approach by considering that the $\Sigma^*(1/2^-)$ couples strongly to the $\bar K N$ channel, and predicted the differential and total cross sections. The total cross section of $\gamma n \to K^+\Sigma^*(1/2^-)$ is approximately 4.2~$\mu$b around $E_\gamma=2.4$~GeV.

Furthermore, similar to photoproduction, Ref.~\cite{Ren:2015bsa} proposed to study the strangeness baryon in the antineutrino induced reactions $\bar{\nu}_l p \to l + \Phi B$ where $\Phi B$ could be various meson-baryon systems with strangeness, for instance, $\bar{K} N$, $\pi \Sigma$, $\pi \Lambda$, and $K\Xi$. They showed that a combined study of $\pi^0\Sigma^0$, $\pi^+\Sigma^-$, and $\pi^-\Sigma^+$ production induced by antineutrinos could provide useful information about the $\Sigma^*(1/2^-)$ state.

\subsection{Charmonium decays $\chi_{c0}\to \bar\Sigma\Sigma \pi, \bar\Lambda\Sigma\pi$, $J/\psi\to \Lambda\bar\Lambda \pi$}

In Ref.~\cite{Wang:2015qta}, the authors proposed to search for the $\Sigma^*(1/2^-)$ in the process $\chi_{c0}(1P)\to \bar{\Sigma}\Sigma\pi$.
Since the quantum numbers of the $\chi_{c0}(1P)$ are $I^G(J^{PC})=0^+(0^{++})$, and the $\chi_{c0}(1P)$, blind to SU(3), behaves like an SU(3) singlet, this process is a good isospin filter that guarantees the $\pi\Sigma$ system has isospin $I=1$. 
On the other hand, the $\chi_{c0}(1P)$ is a SU(3) singlet, and $\bar\Sigma$ belongs to an SU(3) octet. Thus the $\pi\Sigma$ system will also be in an octet state, which suppresses the contribution from the decuplet state $\Sigma(1385)$. 

Considering the $\pi\Sigma$ and $\pi\bar{\Sigma}$ final-state interactions within the chiral unitary approach, Ref.~\cite{Wang:2015qta} predicted one cusp structure around 1430~MeV in the $\pi\Sigma$ invariant mass distribution, which could be associated with the $\Sigma^*(1/2^-)$ predicted in Refs.~\cite{Roca:2013cca,Jido:2003cb,Oller:2000fj}.

Furthermore, Ref.~\cite{Liu:2017hdx} performed a theoretical study on the process $\chi_{c0}\to \bar\Lambda \Sigma \pi$, considering the contributions from the $\pi\Sigma$ and $\pi\bar\Lambda$ final state interactions within the chiral unitary approach, which could dynamically generate the resonance $\Sigma^*(1/2^-)$ with a mass about 1430~MeV and $\Lambda(1405)$. 
Meanwhile, another $\Sigma^*(1/2^-)$ with a mass of 1380~MeV and a width of 120~MeV is considered. 
The predicted $\pi\bar\Lambda$ invariant mass distribution of the process $\chi_{c0}\to \bar\Lambda \Sigma \pi$ has a broad peak around 1380~MeV and a cusp structure around the $\bar{K}N$ threshold, which could be associated with the resonance $\Sigma^*(1/2^-)$~\cite{Liu:2017hdx}. 

Meanwhile, the possible existence of such a $\Sigma^*(1/2^-)$ structure in $J/\psi$ decays was discussed in Ref.~\cite{Zou:2006uh}.
Recently, Ref.~\cite{Huang:2024oai} investigated the isospin-breaking process $J/\psi\to \Lambda\bar\Lambda\pi$, and found that the triangle singularity plays a significant role in this process, resulting in the creation of a resonance-like structure around 1.4~GeV in the $\Lambda\pi$ ($\bar\Lambda\pi$) invariant mass spectrum. 
It is suggested that the process $J/\psi\to \Lambda\bar\Lambda\pi$ could be used to offer direct evidence for the predicted triangle singularity and the additional evidence of the $\Sigma^*(1/2^-)$.

\subsection{Charmed decays $\Lambda_c^+\to \eta\Lambda \pi^+ $}

Based on a 980~fb$^{-1}$ data sample, the Belle Collaboration measured the process $\Lambda^+_c\to \eta\pi^+\Lambda$. Significant signals of the resonances $\Lambda(1670)$ and $\Sigma(1385)$ appear in the $\eta\Lambda$ and $\pi^+\Lambda$ invariant mass distributions~\cite{Belle:2020xku}. 
This process has been measured by CLEO~\cite{CLEO:1995cbq} and BESIII~\cite{BESIII:2018qyg} earlier, and was also suggested to search for the $\Sigma^*(1/2^-)$ in Ref.~\cite{Xie:2017xwx}. 
It is shown that the contribution of $\Sigma^*(1/2^-)$ with a mass about 1380~MeV makes the peak of $\pi^+\Lambda$ invariant mass distribution broader, and the shapes of the angle distributions are very different for $\Sigma(1385)$ and $\Sigma^*(1/2^-)$~\cite{Xie:2017xwx}.

In Ref.~\cite{Wang:2022nac}, the authors analyzed this process by considering the intermediate state $\Sigma^*(1385)$ ($J^P=3/2^+$), and the $S$-wave $\eta\Lambda$ and $\eta\pi^+$ final-state interactions in the chiral unitary approach, which dynamically generate the $\Lambda(1670)$ and $a_0(980)$ resonances. 
Their results could reproduce the Belle data of the $\pi^+\Lambda$ and $\eta\Lambda$ invariant mass distributions. 
However, the Dalitz plot of the process $\Lambda^+_c\to \eta\pi^+\Lambda$ has more yields appearing in the region of $M^2_{\Lambda\pi^+}<1.85$~GeV$^2$ and $M^2_{\eta\Lambda}<4$~GeV$^2$, which cannot be well described in Ref.~\cite{Wang:2022nac}. 
These discrepancies may imply some contributions from the resonance $\Sigma^*(1/2^-)$~\cite{Xie:2017xwx}.

Motivated by the Belle measurements of the process $\Lambda_c^+\to \eta\pi^+\Lambda$~\cite{Kim:2021wov}, Ref.~\cite{Lyu:2024qgc} investigated this process by considering the contributions from the $a_0(980)$, $\Lambda(1670)$, and the intermediate resonances $\Sigma(1385)$ and $\Sigma^*(1/2^-)$. 
It was shown that the results involving the $\Sigma^*(1/2^-)$ are favored by the Belle data of the $\eta\Lambda$ and $\pi^+\Lambda$ invariant mass distributions. 
Furthermore, the $\eta\pi^+$ invariant mass distribution and the angular distribution $d\Gamma/d{\rm cos}\theta$ were predicted, which are significantly different depending on whether the $\Sigma^*(1/2^-)$ is present or not. 
Finally, it was shown that with the contribution from the $\Sigma^*(1/2^-)$, the calculated Dalitz plot agrees with the Belle measurements.

\subsection{Charmed decays $\Lambda_c^+\to p\bar{K}^0\eta$}

Recently, the Belle Collaboration measured the branching fraction $\mathcal{B}(\Lambda_c^+\to pK^0_S \eta)=(4.35\pm 0.10\pm 0.20\pm 0.22)\times 10^{-3}$, and reported the invariant mass squared distributions of this process~\cite{Belle:2022pwd}. 
In the $\eta p$ invariant mass squared distribution, one can find a narrow peak around $M^2_{\eta p}=2.3$~GeV$^2$ and a broad peak around $M^2_{\eta p}=2.75$~GeV$^2$, which should correspond to the $N(1535)$ and $N(1650)$ resonances, respectively~\cite{Xie:2017erh,Pavao:2018wdf}. 
In addition, a threshold enhancement around  $M^2_{ p{K}_S^0}=2.1$~GeV$^2$ in the $p{K}_S^0$ invariant mass squared distribution could be associated with the predicted low-lying excited $\Sigma^*(1/2^-)$ baryon state below the $pK^0_S$ threshold.

Motivated by the Belle measurements on the  process $\Lambda_c^+ \to p \bar{K}^0\eta$, Ref.~\cite{Li:2024rqb} investigated this process by considering the contributions from the $N(1535)$, $N(1650)$, and the predicted low-lying baryon $\Sigma^*(1/2^-)$.

It was demonstrated that with the contribution from $\Sigma^*(1/2^-)$ considered, and the $\chi^2/d.o.f.$ of fit reduces from 5.67 to 1.55, and the calculated invariant mass spectrum agrees with the Belle measurements.
%, as shown by Fig.~\ref{fig:Lc2pKpi}.
%
The near-threshold enhancement in the $p \bar{K}^0$ invariant mass squared distribution could also be well reproduced, which implies that the $\Sigma^*(1/2^-)$ plays an important role in this process~\cite{Li:2024rqb}.

\subsection{Charmed decays $\Lambda_c^+\to \pi^+\pi^0\pi^-\Sigma^+$}

Since triangle singularity plays an important role in many processes~\cite{Liu:2015taa,Bayar:2016ftu,Liang:2019jtr,Dai:2018hqb,Wang:2016dtb,Guo:2019twa},  Ref.~\cite{Xie:2018gbi} studied the $\Sigma^*(1/2^-)$ production in the decay of  $\Lambda_c^+\to \pi^+\pi^0\pi^-\Sigma^+$ by considering the triangle loop, which develops a singularity of the invariant mass of $\pi^0\pi^-\Sigma^+$ system around 1880~MeV.
%, as show in Fig.~\ref{fig:md_Lc23pS}. 
It is found that a cusp structure, tied to the $\Sigma^*(1/2^-)$ state, appears in the final $\pi^-\Sigma^+$ mass spectrum at the energy around the $\bar K N$ threshold as shown in Fig.~6 of Ref.~\cite{Xie:2018gbi}.
It should be pointed out that, the $\Sigma^*(1/2^-)$ with a mass around 1430~MeV could be dynamically generated from the meson-baryon interaction, and appears as a weakly bound state or a virtual state, which is reflected as a strong cusp structure around the $\bar{K}N$ threshold~\cite{Roca:2013cca}.
 Meanwhile, the branching ratio for the $\Sigma^*(1/2^-)$ signal is predicted to be of the order of $10^{-4}$~\cite{Xie:2018gbi}. In the future, Belle II, BESIII, and the future Super Tau-Charm Factory (STCF) experiments could test such predictions.

\section{Summary and outlook}\label{sec:sum}
Although quenched quark models have successfully described the ground-state baryons as three-quark states, there are strong empirical and phenomenological indications for larger sea-quark admixtures in the low-lying excited baryons. 
Investigating the low-lying excited baryon $\Sigma^*(1/2^-)$ is crucial to determining the nature of the low-lying baryons and understanding meson-baryon interactions. 
As the isovector partner of the $\Lambda(1405)$ and $N(1535)$, the $\Sigma^*(1/2^-)$ has motivated many experimental and theoretical works in the last decades.

This review presents the experimental information on the $\Sigma^*(1/2^-)$. 
The $\Sigma(1620)$ is a 1-star resonance in the present RPP, and its existence is still in debate. 
The $\Sigma(1480)$ appeared in the previous version of RPP before 2019, and its existence is still to be confirmed. 
Recently, Belle has observed one cusp structure around the $\bar K N$ threshold in the $\Lambda\pi^+\pi^+\pi^-$ reaction, which could be associated with the predicted $\Sigma^*(1/2^-)$ resonance.

Next, we have provided a brief introduction of the theoretical studies on the $\Sigma^*(1/2^-)$ productions in the processes $K^-p\to \Lambda\pi^+\pi^-$, $\Lambda p\to \Lambda p \pi^0$, $\gamma p \to K^+ \pi^\pm \Sigma^\mp$, which show some indications for the existence of $\Sigma^*(1/2^-)$. Meanwhile, it is shown that the processes of $\chi_{c0}\to \bar\Sigma \Sigma \pi,\bar\Lambda\Sigma\pi$, $J/\psi\to \Lambda\bar\Lambda\pi$, $\Lambda_c^+\to \eta\Lambda\pi^+, p\bar{K}^0\eta$, and $\Lambda_c^+\to \pi^+\pi^0\pi^-\Sigma^+$ could be used to search for the  $\Sigma^*(1/2^-)$ in the future.

By now, the BESIII, Belle, LHCb, and J-PARC Collaborations have accumulated a wealth of data about the low-lying excited baryons. In the future, BESIII~\cite{BESIII:2020nme}, Belle II~\cite{Jia:2023upb}, LHCb~\cite{LHCb:2022sck}, and J-PARC~\cite{Agari:2012kid} experiments, also the future STCF~\cite{Achasov:2023gey}, are expected to yield more data on the relevant processes, which could shed light on the existence of the low-lying excited $\Sigma^*(1/2^-)$ state.

Another possible ideal clean platform to study the low-lying excited $\Sigma^*(1/2^-)$ state is the $\bar{\nu}_{l}+N \to l^+ + \Sigma^* \to  l^+ + \Lambda\pi$ reactions~\cite{Wu:2013kla} to avoid complicated strong final-state interactions with other hadrons in electric or hadron production processes. 
With increasing intensity of neutrino fluxes, there are many new measurements
of neutrino nuclear scattering from various groups at Fermilab~\cite{ArgoNeuT:2014rlj}, KEK~\cite{T2K:2014hih} and CERN~\cite{FASER:2023zcr}. The study of $\bar{\nu}_{l}+N \to l^+ + \Sigma^* \to  l^+ + \Lambda\pi$ reaction may be considered and realized.

\section*{Acknowledgements}
We would like to thank Eulogio Oset for useful comments.
LSG and JJX are partly supported by the National Key R\&D Program of China under Grant No. 2023YFA1606700. EW is partly supported by the National Key R\&D Program of China under Grant No. 2024YFE0105200.
This work is supported by the Natural Science Foundation of Henan under Grant No. 232300421140 and No. 222300420554, the National Natural Science Foundation of China under Grant Nos. 12475086, 12192263, 12205075, 12175239, 12221005, 12075288, 12361141819, 
and also by the Open Project of Guangxi Key Laboratory of Nuclear Physics and Nuclear Technology, No. NLK2021-08,
and also by the Central Government Guidance Funds for Local Scientific and Technological Development, China (No. Guike ZY22096024), 
and also by the National Key Research and Development Program of China under Contracts 2020YFA0406400, 
and also by the Chinese Academy of Sciences under Grant No. YSBR-101, 
and also by the Youth Innovation Promotion Association CAS.

\end{document}